\documentclass[a4paper,12pt]{article}
\pdfoutput=1

\usepackage{jheppub}
\usepackage[T1]{fontenc}
\usepackage{amsmath}
\usepackage{amsfonts}
\usepackage{amssymb}
\usepackage{color}
\usepackage{graphicx}
\usepackage{mathrsfs}
\usepackage{graphics}
\usepackage{epstopdf}

\title{P-T phase diagram of a holographic s+p model from Gauss-Bonnet gravity} \vskip 2cm \vskip 2cm
\author[a,b,1]{Zhang-Yu Nie,\note{Corresponding author.}}
\author[a,b]{Hui Zeng}

\affiliation[a]{Kunming University of Science and Technology, Kunming 650500, China}
\affiliation[b]{State Key Laboratory of Theoretical Physics, Institute of Theoretical Physics, Chinese Academy of Sciences, P.O.Box 2735, Beijing 100190, China}

\emailAdd{niezy@kmust.edu.cn}
\emailAdd{zenghui@kmust.edu.cn}

\abstract{In this paper, we study the holographic s+p model in 5-dimensional bulk gravity with the Gauss-Bonnet term. We work in the probe limit and give the $\Delta$-T phase diagrams at three different values of the Gauss-Bonnet coefficient to show the effect of the Gauss-Bonnet term. We also construct the P-T phase diagrams for the holographic system using two different definitions of the pressure and compare the results.}
\begin{document}
\maketitle
\flushbottom

\section{\bf Introduction}
The AdS/CFT correspondence~\cite{Maldacena:1997re,Gubser:1998bc,Witten:1998qj} has been widely studied over the past years. As a strong-weak duality, it is believed to be a useful tool to study the strongly coupled systems in QCD~\cite{Sakai:2004cn,Karch:2006pv} as well as condensed matter physics. Although it is still far from the practical application, some qualitative key features has been captured, such as the phase transition behavior and optical conductivity in holographic superconductors~\cite{Gubser:2008px,Hartnoll:2008vx,Herzog:2008he}. Recent studies have extended the holographic models to non homogenous systems\cite{Donos:2011bh}, time dependent evolutions\cite{Chesler:2008hg,Bhattacharyya:2009uu,Murata:2010dx,Bhaseen:2012gg} as well as systems with multiple order parameters. More and more phenomenons in real systems can be realized holographically at present.

Some physical system exhibit complex phase structures, and need to be described in models with multiple order parameters. Holographic models have also been setup to study the competition and coexistence effect between the different orders in such systems. In Ref.~\cite{Basu:2010fa} the authors considered the competition effect of two scalar order parameters in the probe limit and found the signal of a coexisting phase where both the two s-wave order parameters have non zero value.  In Ref.~\cite{Cai:2013wma}, the authors considered two scalars charged under the same U(1) gauge field with full back reaction of matter fields on the background geometry.  It turns out that the model has a rich phase structure for the competition and coexistence of the two scalar orders. In Ref.~\cite{Nie:2013sda,Nie:2014qma}, the authors studied the competition and coexistence between the s-wave and p-wave orders. They found that the orders with different symmetry can also coexist in some situation and found that the back reaction of the matter fields on the back ground geometry can open the door to novel phase transition behaviors and rich phase structures. Further studies also extended the holographic study of competing orders to s+d systems~\cite{Nishida:2014lta,Li:2014wca}. More work concerning the competition of multi order parameters can be found in Refs.~\cite{Huang:2011ac,DG,Krikun:2012yj,Donos:2012yu,AAJL,Musso:2013ija,Nitti:2013xaa,Liu:2013yaa,Amado:2013lia,Amoretti:2013oia,Momeni:2013bca,Donos:2013woa,Chaturvedi:2014dga,Giordano:2015vsa,Nishida:2015ipa,Liu:2015zca}, and see Ref.~\cite{Cai:2015cya} for a recent review on holographic superconductor models.

\begin{figure}\center
\includegraphics[width=10.6cm] {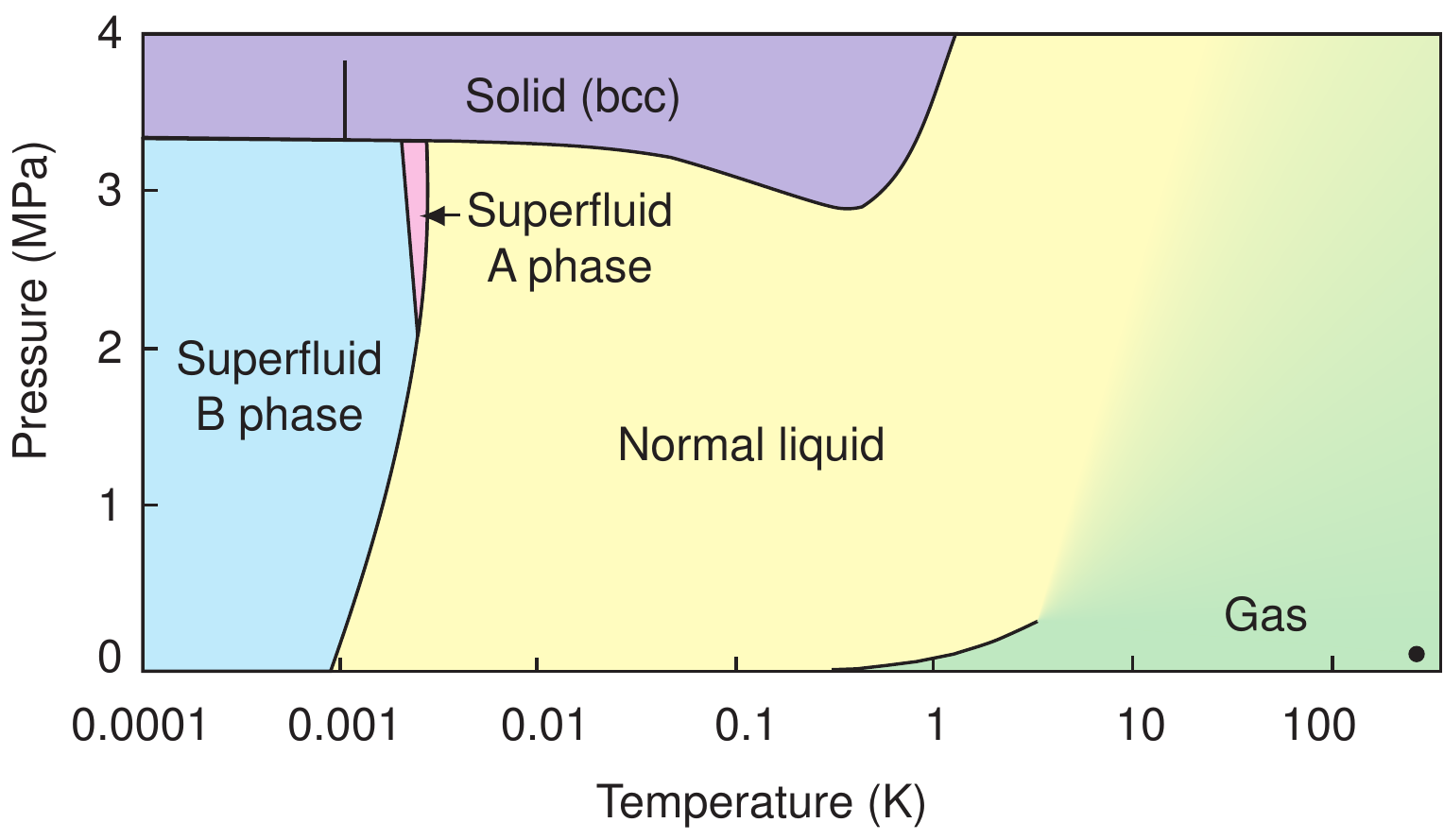}
\caption{\label{phasediagram1}P-T phase diagram for Helium-3.}
\end{figure}

In real world, there are also systems contain multi-order parameters and exhibit competition behavior in the phase structure. The superfluid Helium-3 is a well known material that suffers superfluid phase transitions at very low temperature~\cite{Volhardt-Wolfle-1990}. Figure~\ref{phasediagram1} is a P-T phase diagram\footnote{Colored figure from: http://ltl.tkk.fi/research/theory/helium.html} of Helium-3. We can see in this figure that there are two different superfluid phases of liquid Helium-3 labeled "A-phase" and "B-phase" respectively. The A-phase is the anisotropic p-wave superfluid phase, and the B-phase is the isotropic p+$i$p superfluid phase. It seems that the isotropic phase is favored at a lower temperature and lower pressure, while the anisotropic phase dominate the region with higher temperature and larger pressure. The holographic s+p system also contains an isotropic phase and an anisotropic one. Therefore it would be interesting to study the P-T phase diagram of the holographic s+p system to see whether there are some universalities between the isotropic and anisotropic superfluid phases in the P-T phase diagram.

Some recent studies have already realized P-T phase diagrams in gravitational systems. The authors of Ref.~\cite{Kubiznak:2012wp} have studied charged AdS black hole in the extended phase space, where the cosmological constant appears as the thermodynamic pressure and its conjugate quantity as a thermodynamic volume of the black hole~\cite{Kastor:2009wy}. The gravitational system in the extended phase space share the same critical exponents with a van der Waals system, and they have extremely similar diagrams in the P-V plane. This study has also been extended to the Einstein-Gauss-Bonnet gravity~\cite{Cai:2013qga}, where the relation between the pressure and cosmological constant is not affected by the Gauss-Bonnet term. 

There is also another definition for the pressure, which can be deduced from the stress-energy tensor of the boundary conformal fluid. This definition of pressure is different from the above one, and is more meaningful from the view point of the boundary dual CFT. Therefore, in this paper, we will construct the P-T phase diagrams using the two different ways of defining the pressure respectively, and compare the results.

This article is organized as follows. In Sec.~\ref{sect:setup} we give our model set up and shortly introduce the necessary calculations. In Sec.~\ref{sect:deltaT}, we draw the $\Delta-\text{T}$ phase diagrams of this model in different values of Gauss-Bonnet coefficient to see the effect of the Gauss-Bonnet term on the competition and coexistence of the s-wave and p-wave orders. We use the two different definitions of pressure to construct the P-T phase diagrams and compare the results in Sec.~\ref{sect:PT}. Our conclusions and discussions are included in Sec.~\ref{sect:conclusion}.

\section{Holographic model of an s+p superconductor}
\label{sect:setup}
\subsection{The model setup}
We still extend the $SU(2)$ holographic p-wave model to get an s+p model for simplicity. The results in this paper can also be get form the extension of the complex vector holographic p-wave model~\cite{Cai:2013aca,Wu:2014bba} as given in Ref.~\cite{Nie:2014qma}. The action of the holographic s+p model in Einstein-Gauss-Bonnet gravity is 
\begin{eqnarray}
S  &=&S_G+S_M,\\
S_G&=&\frac{1}{2 \kappa_g ^2}\int d^{5}x \sqrt{-g} \big(R-2\Lambda+\frac{\alpha}{2}(R^2-4R^{\mu\nu}R_{\mu\nu}+R^{\mu\nu\rho\sigma}R_{\mu\nu\rho\sigma})\big),\\
S_M&=&\frac{1}{g_c^2}\int d^{5}x \sqrt{-g}(-D_\mu \Psi^{a} D^\mu \Psi^a-\frac{1}{4}F^a_{\mu\nu}F^{a\mu\nu}-m^2 \Psi^a\Psi^a). \label{Smatter}
\end{eqnarray}
Here $\Psi^a$ is the SU(2) charged scalar triplet in the vector representation of the SU(2) gauge group, and
\begin{equation}
D_\mu\Psi^a=\partial_\mu \Psi^a+\varepsilon^{abc}A^b_\mu\Psi^c.
\end{equation}
$F^a_{\mu\nu}$ is the gauge field strength, and is given by
\begin{equation}
F^a_{\mu\nu}=\partial_\mu A^a_\nu-\partial_\nu A^a_\mu +\varepsilon^{abc} A^b_\mu A^c_\nu.
\end{equation}
$g_c$ is the Yang-Mills coupling constant as well as the SU(2) charge of $\Psi^a$.

The gravitational action with Gauss-Bonnet term admits an analytical black brane solution given by~\cite{Cai:2001dz}
\begin{eqnarray}\label{metric}
ds^2&=&-f(r)dt^2+\frac{1}{f(r)}dr^2+r^2dx^{2}_i,
\end{eqnarray}
where $x^i$s are the coordinates of a three dimensional Euclidean space with $i\in\{1,2,3\}$ and
\begin{eqnarray}
f(r)=\frac{r^2}{2\alpha}\Big(1-\sqrt{1-\frac{4\alpha}{L^2}(1-\frac{r_h^4}{r^4})} ~\Big).
\end{eqnarray}
This solution is asymptotically AdS with an effective AdS radius
\begin{equation}\label{Leff}
L_{\text{eff}}^2=\frac{1+\sqrt{1-\frac{4\alpha}{L^2}}}{2} L^2.
\end{equation}
The temperature of this solution is
\begin{equation}\label{Temperature}
T=\frac{1}{\pi L^2} r_h.
\end{equation}

To solve the s+p system consistently, we should in principle solve the equations of motion for the matter fields and that for the metric fields together. However, in this paper, we will study the system in the probe limit. This means that the back reaction of the matter fields on the background geometry is neglected. This limit can be realized consistently by taking the limit $g_c \gg \kappa_g$. In the probe limit, we can only consider the equations of motion for the matter fields on the above gravitational space time background.

We consider the following ansatz  for the matter fields
\begin{eqnarray}
\Psi^3=\Psi_3(r),~A^1_t=\phi(r),~A^3_x=\Psi_x(r),
\end{eqnarray}
with all other field components being turned off. In this ansatz, we take $A^1_\mu$ as the electromagnetic U(1) field as in Ref.~\cite{Gubser:2008wv}. With this ansatz, the equations of motion  of matter fields in the AdS black brane background read
\begin{eqnarray}
\phi''+\frac{3}{r}\phi' -(\frac{2  \Psi_3^2}{f}+\frac{\Psi_x^2}{r^2 f})\phi&=&0, \label{eqphi}\\
\Psi_x''+(\frac{1}{r}+\frac{f'}{f})\Psi_x'+\frac{\phi^2}{f^2}\Psi_x&=&0, \label{eqpsix}\\
\Psi_3''+(\frac{3}{r}+\frac{f'}{f})\Psi_3'-(\frac{m^2}{f}-\frac{\phi^2}{f^2})\Psi_3&=&0. \label{eqpsi3}
\end{eqnarray}
We can see $\Psi^3$ and $\Psi_x$ are not directly coupled in their equations of motion, and they both coupled to the same U(1) electromagnetic field. Thus we can consistently set $\Psi_x=0$ to get the same equations for the s-wave holographic superconductor model~\cite{Pan:2009xa}, or set $\Psi_3=0$ to get the equations for the $SU(2)$ p-wave holographic superconductor model~\cite{Cai:2010cv}. We can further find solutions dual to an s+p coexistent phase with both $\Psi_x$ and $\Psi_3$ non-zero within some regions of the parameters $m^2$ and $\alpha$.

\subsection{Boundary conditions}
To solve the equations of motion \eqref{eqphi},\eqref{eqpsix},\eqref{eqpsi3}, we need to specify the boundary conditions both on the horizon and on the boundary. Our choice of the boundary conditions are the same as that in the individual s-wave and p-wave holographic superconductor models. We set the source term of the s-wave and p-wave order to be zero and get the spontaneously symmetry broken solutions.

The boundary behaviors of the three fields on the horizon side are
\begin{eqnarray}\label{shorizon}
\phi(r)&=& \phi_1(r-r_h)+\mathcal{O}((r-r_h)^2),\nonumber \\
\Psi_x(r)&=& \Psi_{x0} + \mathcal{O}(r-r_h),\nonumber \\
\Psi_3(r)&=& \Psi_{30} + \mathcal{O}(r-r_h).
\end{eqnarray}

Since $\phi(r)$ is the $t$ component of the vector field $A_\mu^1$, so $\phi(r_h)$ is set to zero to avoid the divergence of $g^{\mu\nu}A_\mu^1 A_\nu^1$ at the horizon. And Eq.(\ref{eqpsix},\ref{eqpsi3}) impose constraints on the derivative of $\Psi_x(r)$ and $\Psi_3(r)$ at the horizon. So the expansions of the functions $\phi(r)$, $\Psi_x(r)$ and $\Psi_3(r)$ near the horizon have only three free parameters. In other words, the solutions to the equations of motion (\ref{eqphi},\ref{eqpsix},\ref{eqpsi3}) are determined by the three parameters $\phi_1$, $\Psi_{x0}$ and $\Psi_{30}$.

The expansions of $\phi(r)$, $\Psi_x(r)$ and $\Psi_3(r)$ near the AdS boundary are of the forms
\begin{eqnarray}\label{sboundary}
\phi(r)&=& \mu-\frac{\rho}{r^{2}}+...~, \nonumber\\
\Psi_x(r)&=& \Psi_{xs} + \frac{\Psi_{xe}}{r^{2}}+... ~, \nonumber\\
\Psi_3(r)&=& \frac{\Psi_-}{r^{\Delta_-}} + \frac{\Psi_+}{r^{\Delta_+}}+... ~,
\end{eqnarray}
where $\mu, \rho$ are the chemical potential and the charge density, and~\cite{Pan:2009xa}
\begin{equation}\label{Deltapm}
\Delta_\pm=2\pm \sqrt{4+m^2 L_{\text{eff}}^2}.
\end{equation}
We will choose $\Psi_{xs}$ and $\Psi_-$ as the source terms of the p-wave and s-wave operators respectively, therefore $\Psi_{xe}$ and $\Psi_+$ are the corresponding expectation values. Thus the conformal dimension of the s-wave order is 
\begin{equation}\label{Delta}
\Delta\equiv\Delta_+=2+ \sqrt{4+m^2 L_{\text{eff}}^2}.
\end{equation}

In the action of this model, we have two free parameters $m^2$ and $\alpha$. In the horizon expansion of the three functions $\phi(r)$, $\Psi_x(r)$ and $\Psi_3(r)$, we have additional 3 free parameters. In order to get the solutions with spontaneous charge condensation, we should set two constraints $\Psi_{xs}=0, \Psi_-=0$ on the boundary. As a results, we have 3 free parameters left to fix the solution. These free parameters can be chosen as $\{\Delta,\alpha,T\}$ according to the relations (\ref{Temperature}),(\ref{Delta}).

\subsection{Different solutions and the free energy}
With the boundary conditions in the above subsection, we can find solutions at different values of $\{\Delta,\alpha,T\}$. At each point with fixed values of the three parameters, we can easily find a trivial solution with all the matter fields equal to zero. Besides this trivial solution, there are four kinds of solutions dual to different phases on the boundary field theory. We label the four different solutions as N, S, P and SP respectively, and list their properties and the dual phases in boundary field theory in Table~\ref{Table1}.

\begin{table}
\center{}
\begin{tabular}{|c|c|c|}
  \hline
    Label of the solution & Properties of the solution  & Related Phase in CFT				\\\hline
    Solution-N &   $\phi(r)\neq 0,\Psi_x(r)=0,\Psi_3(r)=0$  &  Normal phase	\\\hline
    Solution-S &   $\phi(r)\neq 0,\Psi_x(r)=0,\Psi_3(r)\neq 0$   &  s-wave superfluid phase \\\hline
    Solution-P &   $\phi(r)\neq 0,\Psi_x(r)\neq 0,\Psi_3(r)=0$   &  p-wave superfluid phase \\\hline
    Solution-SP &   $\phi(r)\neq 0,\Psi_x(r)\neq 0,\Psi_3(r)\neq 0$   & s+p superfluid phase\\
  \hline
\end{tabular}
\caption{\label{Table1}The different solutions. }
\end{table}

Solution-N is dual to the normal phase in the boundary theory, and exist in every point in the parameter space spanned by $\{\Delta,\alpha,T\}$. The other solutions are dual to superfluid phases that exist in different regions of the parameter space. These four solutions would overlap in some region, and in such region,  we should compare the free energy for all the possible solutions to determine which one is the most stable. We give the formula for calculating the free energy of different solutions below. 

We work in the grand canonical ensemble with fixed chemical potential and calculate the Gibbs free energy for the four solutions. The Gibbs free energy of the system can be identified with the temperature times the Euclidean on-shell action of the bulk solution. Because we work in the probe limit, the differences between the free energies only come from the matter part of the action. Thus we need only calculate the contribution of matter fields to the free energy:
\begin{equation}
\Omega_m=T S_{ME},
\end{equation}
where $S_{ME}$  denotes the Euclidean on-shell action of matter fields in the black brane background. Note that since we work in the grand canonical ensemble and choose the scalar operator with dimension $\Delta_+$,  no additional surface term and counter term for the matter fields are needed. It turns out that the Gibbs free energy can be expressed as
\begin{equation}
\Omega_m=\frac{V_3}{g_c^2} (-\mu\rho + \int_{rh}^\infty (\frac{r \phi^2\Psi_x^2}{2f}+\frac{r^3 \phi^2\Psi_3^2}{f})dr).
\end{equation}
Here $V_3$ denotes the area of the 3-dimensional transverse space.

We have calculated the Gibbs free energy for the four different phases, and confirmed that the Solution-SP is always stable once it exist. The other three phases are partly stable. With the information of the free energies for different phases, we can construct a three dimensional $\Delta-\alpha-T$ phase diagram for this holographic system. We can otherwise fix the value of one parameter and get a two dimensional phase diagram for simplicity.

\section{$\Delta-\text{T}$ phase diagrams of the s+p model in varies Gauss-Bonnet coefficient}
\label{sect:deltaT}
In this section, we fix $\alpha$ to three different values $\alpha/L^2=0.0001, 0.09, -0.19$, and draw the $\Delta-\text{T}$ phase diagrams to show the effect of Gauss-Bonnet term. We choose these three special values because that there is a constrain to the Gauss-Bonnet coefficient as $-7/36<\alpha/L^2<9/100$ from the boundary causality~\cite{Brigante:2008gz,Buchel:2009tt}, and the $\alpha=0.0001 L^2$ phase diagram is very close to that with $\alpha=0$. Note that in order to compare the holographic system in different values of the Gauss-Bonnet coefficient, we have set $L_{\text{eff}}=1$. We will give the reason in the next section.

\begin{figure}
\includegraphics[width=7.5cm] {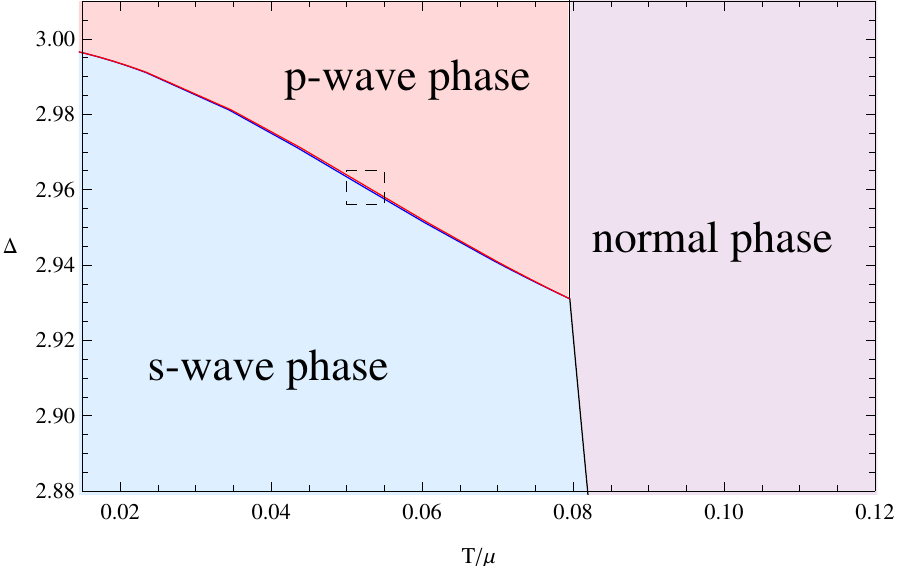}
\includegraphics[width=7.5cm] {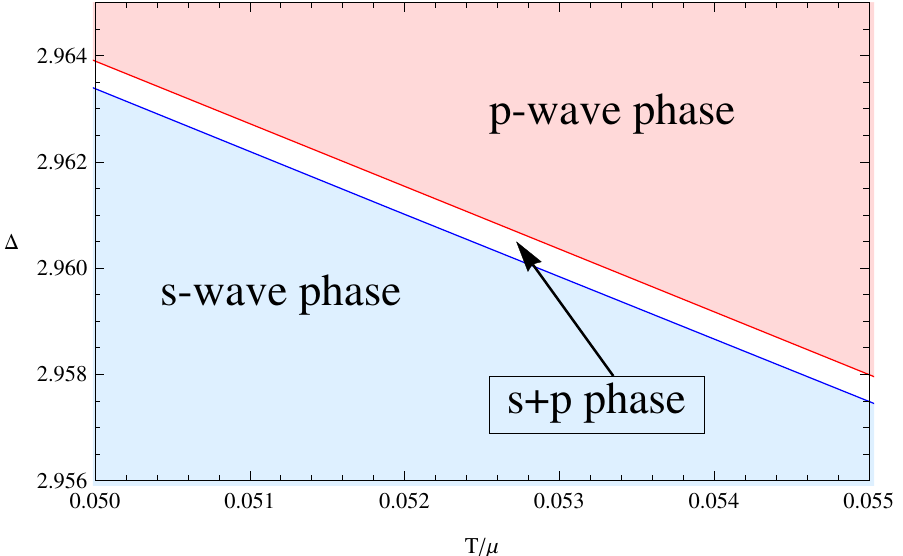}
\caption{\label{deltaT1}The $\Delta-\text{T}$ phase diagram with $\alpha=0.0001 L^2$. The right figure is and enlarged version for the left one. We use different colors to denote the different stable phases. The light purple denotes the normal phase, light red and light blue denote the p-wave and s-wave phases respectively. The region for the s+p phase is colored white, and only can be seen clearly in the enlarged figure. The curves separating the different phases are made up by critical points.}
\end{figure}

We give the $\Delta-\text{T}$ phase diagram with $\alpha=0.0001 L^2$ in Figure~\ref{deltaT1}, where the right one is an enlarged version of the dashed rectangle region in the left one. We can see that in this phase diagram there are four regions marked with different colors. The light purple region denote the normal phase where both the s-wave and p-wave orders do not condense. The light red region denote the p-wave superfluid phase, and the light blue region denote the s-wave superfluid phase. In these two cases, only one order of the two condense. The fourth region marked white denote the s+p phase, where both the s-wave and p-wave orders condense. This white region is between the light red and light blue regions, but it is too narrow to be seen clearly in the left figure. So we draw an enlarged version of the dashed rectangle region in the left figure on right to show more details. We can see in the enlarged figure that the s+p phase indeed exist.

\begin{figure}
\includegraphics[width=7.5cm] {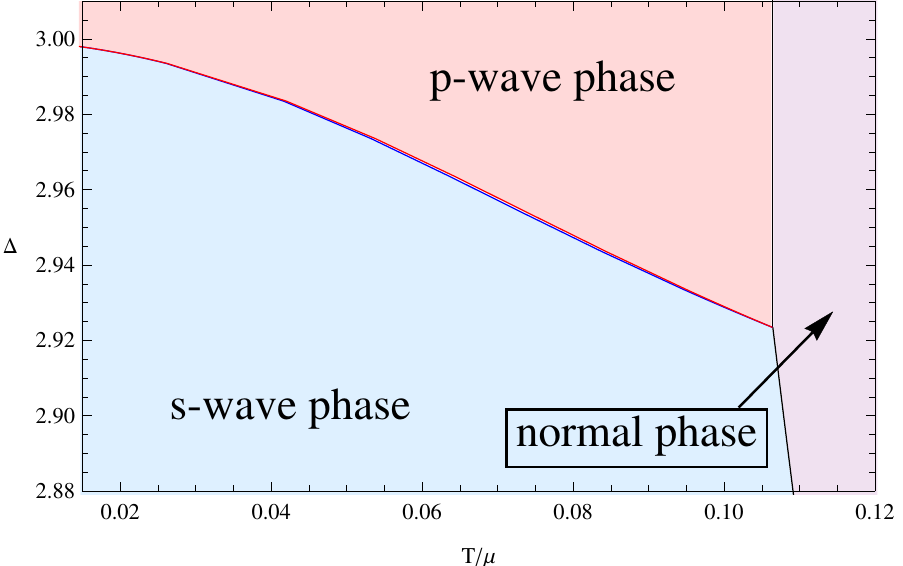}
\includegraphics[width=7.5cm] {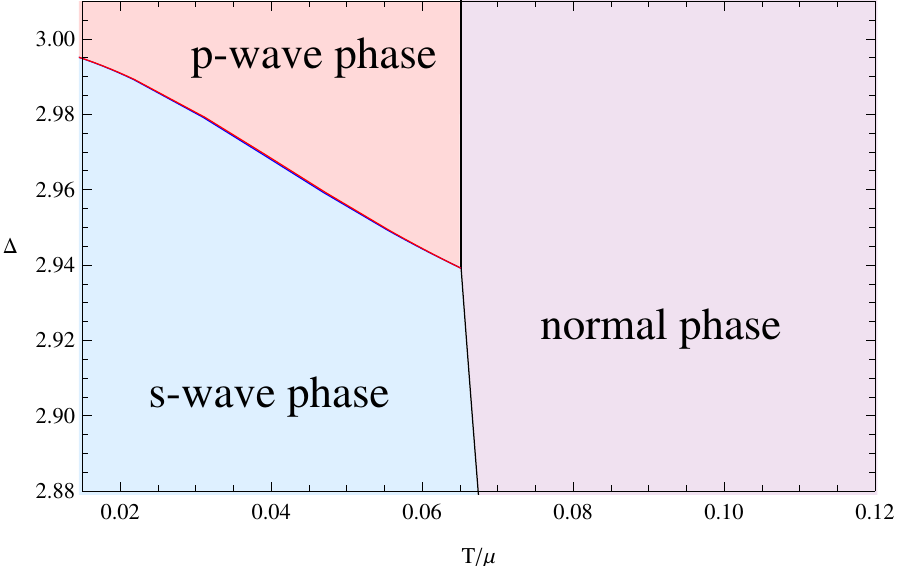}
\caption{\label{deltaT2}The $\Delta-\text{T}$ phase diagram with $\alpha=-0.19 L^2$(Left) and $\alpha=0.09 L^2$(Right). The notions in these figures are the same as that in Figure~\ref{deltaT1}.}
\end{figure}

The $\Delta-\text{T}$ phase diagrams with $\alpha=-0.19 L^2$ and $\alpha =0.09 L^2$ are given in Figure~\ref{deltaT2}. We can see these two phase diagrams are qualitatively the same as that with $\alpha=0.0001 L^2$. So the effect of Gauss-Bonnet term on the phase diagram is rather simple. We can see clearly that when the Gauss-Bonnet coefficient is larger, the critical temperatures of phase transitions from the normal phase are lower, and the the value of $\Delta$ for the quadruple intersection point is larger.

The $\alpha=0.0001 L^2$ phase diagram is very close to the one with $\alpha=0$. Thus we can compare this diagram to the previous results of this model in 4-D Einstein gravity~\cite{Nie:2013sda,Nie:2014qma}. Actually, we can see all the three phase diagrams with different values of $\alpha$ are similar, and they are qualitatively the same as the one from 4-D bulk in probe limit~\cite{Nie:2013sda}. Thus we can see that unlike the effect of turning on the back reaction of the matter fields on the bulk metric~\cite{Nie:2014qma}, the spacetime dimension of the bulk as well as the Gauss-Bonnet term do not bring qualitative change in this holographic s+p model.

\section{P-T phase diagrams from Einstein-Gauss-Bonnet gravity}
\label{sect:PT}
In the previous section, we have shown the $\Delta-\text{T}$ phase diagrams in different values of $\alpha$. In this section, we will introduce our strategy to get the P-T phase digram for the holographic s+p model.

In order to get the P-T phase diagram, we should first introduce the definition of pressure in the gravitational system. There are two different ways to define the pressure in the study of asymptotically AdS black holes. One is to define the pressure as the cosmological constant and study bulk space time in extended phase space, while the other is to define the pressure from the boundary stress energy tensor. In the following two subsections, we use the two different ways to define the pressure and construct P-T phase diagrams for the s+p model in Gauss-Bonnet gravity.

\subsection{Pressure defined by the cosmological constant}

The pressure of a $d$-dimensional gravity system can be related to the cosmological constant in the geometric units as
\begin{equation}\label{pressure}
\text{P}=-\frac{1}{8\pi} \Lambda=\frac{(d-1)(d-2)}{16\pi L^2},
\end{equation}
and the conjugate quantity is the thermodynamic volume of the system. With this treatment, one can reconsider the critical behavior of AdS black holes in an extended phase space, including pressure and volume as thermodynamic variables. In this way, the authors of Ref.~\cite{Kubiznak:2012wp} initiated the investigation of the P-V critical behavior of a charged AdS black hole, and found the same critical behavior and P-V diagram to that for the van der Waals liquid-gas system. The authors of Ref.~\cite{Cai:2013qga} extended the study to Gauss-Bonnet gravity, and find that the relation (\ref{pressure}) holds in  Gauss-Bonnet gravity.

In this paper, we only work in the probe limit and fix the background geometry. As a result, we can not study the complete extended phase space as in Refs.~\cite{Kubiznak:2012wp,Cai:2013qga}. But we can still borrow the relation (\ref{pressure}) and get a P-T phase diagram with the variation of the Gauss-Bonnet coefficient.

In this paper, we have taken $d=5$. The effective AdS radius $L_{\text{eff}}$ is corrected to Eq. (\ref{Leff}). In the AdS/CFT correspondence, the AdS radius is related to the microscopic scale of the dual boundary theory. Therefore, in order to compare the holographic systems with different values of $\alpha$, the effective AdS radius $L_{\text{eff}}$ should be set to the same value. Thus we set $L_{\text{eff}}=1$ instead of the usual choice $L=1$. As a result, the value of $L$, and thus the pressure differs at different values of $\alpha$. We can substitute (\ref{Leff}) to (\ref{pressure}) and get
\begin{equation}\label{P-alpha}
\text{P}=\frac{3}{8\pi L_{\text{eff}}^2} (1+\sqrt{1-4\bar{\alpha}})=\frac{3}{8\pi} (1+\sqrt{1-4\bar{\alpha}}),
\end{equation}
where $\bar{\alpha}=\alpha/L^2$ is a dimensionless parameter. We can use the above relation to translate the change of $\bar{\alpha}$ to the change of pressure P, and the three free parameters of the phase space can now be chosen as $\{\Delta,\text{P},\text{T}\}$. We can draw P-T phase diagrams at different values of $\Delta$.

\begin{figure}
\includegraphics[width=7.5cm] {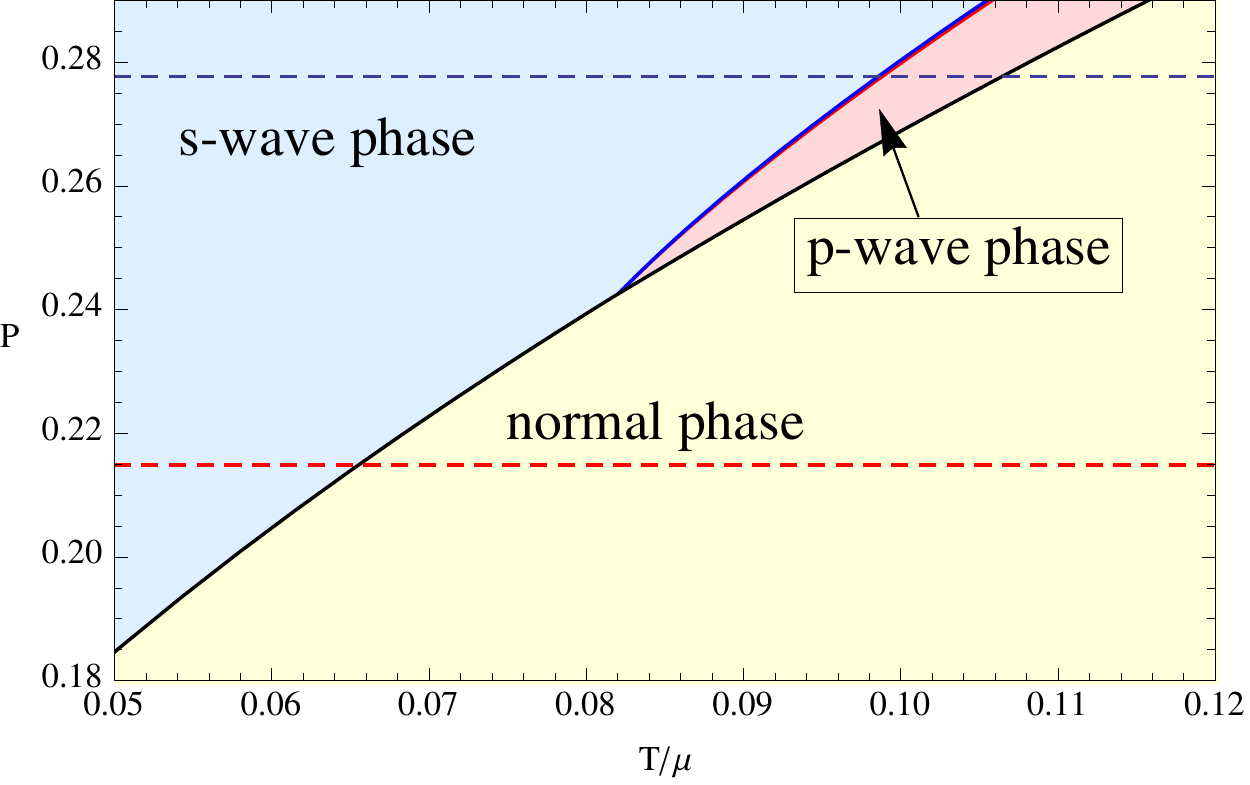}
\includegraphics[width=7.5cm] {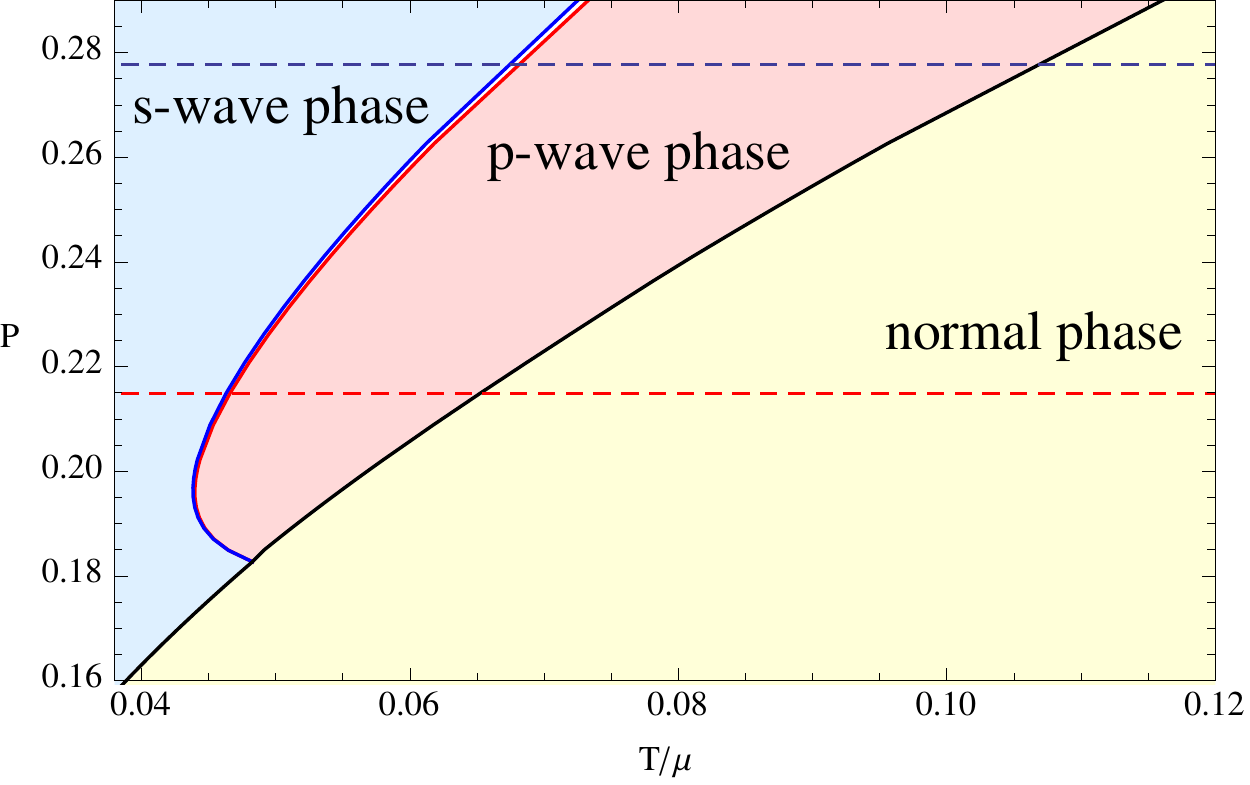}
\caption{\label{PTphasediagram}Two typical P-T phase diagrams with $\Delta=2.93$(Left) and $\Delta=2.96$(Right). We use light yellow to denote the region for the normal phase, while use light red and light blue to denote the p-wave and s-wave superfluid phases respectively. The region for the s+p phase is colored white, but it is too narrow to be seen clearly in these two figure. The dashed blue line is $\alpha/L^2=-7/36$ and the dashed red line is $\alpha/L^2=9/100$.}
\end{figure}

We show two typical P-T phase diagrams with $\Delta=2.93$ and $\Delta=2.96$ in Figure~\ref{PTphasediagram}. We use light yellow to denote the region for the normal phase, while use light red and light blue to denote the p-wave and s-wave superfluid phases respectively. The region for the s+p phase is colored white, however it is too narrow to be seen clearly in the figures. The two dashed horizontal lines are get from the causality constraints $\alpha/L^2>-7/36$(blue) and $\alpha/L^2<9/100$(red) via (\ref{P-alpha}).

We can see from Figure~\ref{PTphasediagram} that the slope of critical line between the normal phase and the superfluid phases is positive. Moreover, the the p-wave phase is favored in the region with higher temperature and pressure, while the isotropic s-wave phase dominate the region with lower temperature and pressure. These qualitative features are very similar to that of the superfluid Helium-3, where the anisotropic p-wave phase is favored in the region with higher temperature and pressure, and the isotropic p+ip phase dominate the region with lower temperature and pressure.

However, this definition of pressure is based on considering the whole bulk spacetime as a thermodynamic system. We can check that the unit of this pressure is that in a 4+1 dimensional spacetime. Therefore the phase diagrams in Figure~\ref{PTphasediagram} are only meaningful for the gravitational system. However, remember that the s-wave phase and p-wave phase are concepts from the conformal field theory, thus in this case, physical meanings of the s-wave phase and p-wave phase in the P-T phase diagram for a gravitational system are not clear.

\subsection{Pressure defined by the energy momentum tensor}
We have successfully got the  P-T phase diagram in the above subsection. However, we should notice that the pressure in the above subsection is defined in the 4+1 dimensional gravitational system.  We can check the dimension of this pressure to confirm this. Therefore the phase diagrams in Figure~\ref{PTphasediagram} are only meaningful for the gravitational system, but not for the boundary CFT. A solution to this problem is to define the pressure from the boundary CFT point of view.

In the AdS/CFT correspondence, the pressure of the conformal field theory can be calculated from the boundary stress energy tensor in asymptotically AdS spacetime. In 5-dimensional Einstein-Gauss-Bonnet gravity with negative cosmological constant, the boundary stress energy tensor can be calculated by the variation of the total action(including the Gibbs-Hawking term and boundary counter terms) with respect to the boundary induced metric~\cite{Balasubramanian:1999re,Myers:1999psa,Bianchi:2001kw,Brihaye:2008xu,Hu:2010sn}
\begin{eqnarray}
T^{ab} &=& \frac{2}{\sqrt{-\gamma}}\frac{\delta S_{total}}{\delta \gamma_{ab}} \nonumber\\
&=&\frac{1}{8\pi G}[K^{ab}-\gamma^{ab}K+\alpha (Q^{ab}-\frac{1}{3}Q \gamma^{ab}) \nonumber\\
&&-\frac{2+U}{L_{\text{eff}}}\gamma^{ab} +\frac{L_{\text{eff}}}{2}(2-U)(\bar{R}^{ab}-\frac{1}{2}\gamma^{ab}\bar{R})],
\end{eqnarray}
where $K_{ab}$ is the extrinsic curvature of the boundary, $\bar{R}_{ab}$ is the Ricci tensor calculated by the boundary induced metric $\gamma_{ab}$, and
\begin{eqnarray}
&U=\sqrt{1-\frac{4\alpha}{L^2}},\\
Q_{ab}=&2K K_{ac}K^c_b -2K_{ac}K^{cd}K_{db} +K_{ab}(K_{cd}K^{cd}-K^2) \nonumber\\
&+2K \bar{R}_{ab}+\bar{R}K_{ab}-2K^{cd}\bar{R}_{cadb}-4\bar{R}_{ac}K_b^c.
&
\end{eqnarray}
The stress energy tensor $\tau^{ab}$ of the conformal field theory dual to the asymptotically AdS bulk can be obtained through the following relation~\cite{Myers:1999psa}
\begin{eqnarray}
\sqrt{-h}h^{ab}\tau_{ab}=\sqrt{-\gamma}\gamma^{ab}T_{ab},
\end{eqnarray}
where $h_{ab}$ is the background metric on which the dual conformal field theory live. We can choose the metric $h_{ab}$ to be the Minkowski metric
\begin{eqnarray}
\eta_{ab}dx^a dx^b=-dt^2+dx_i^2,
\end{eqnarray}
therefore the final expression for the energy momentum tensor for the conformal field theory dual to the spacetime(\ref{metric}) can be written in the form of a perfect fluid as
\begin{eqnarray}
\tau_{ab}=\frac{r_h^4}{16\pi G L_{\text{eff}}^3 L^2} (\eta_{ab}+4u_a u_b).
\end{eqnarray}
We can get the formula for the pressure of the dual CFT from the spacial components of the stress energy tensor. In this paper, we work in probe limit, which means the bulk metric would not be affected by the matter fields. As a result, we can always get an isotropic pressure in different phases as
\begin{eqnarray}\label{pressure2}
P=\frac{r_h^4}{16\pi G L_{\text{eff}}^3 L^2}=\frac{r_h^4}{32\pi G L_{\text{eff}}^5}(1+\sqrt{1-4\bar{\alpha}}) .
\end{eqnarray}
This definition of pressure is standard in the study of AdS/CFT, and is different from that in (\ref{pressure}). One can easily check that the dimension of this pressure is consistent with that in fluid system living on 3+1 spacetime. Therefore, if we want to study the phase structure of the boundary CFT, we would better use the definition in this subsection.

\begin{figure}
\includegraphics[width=7.5cm] {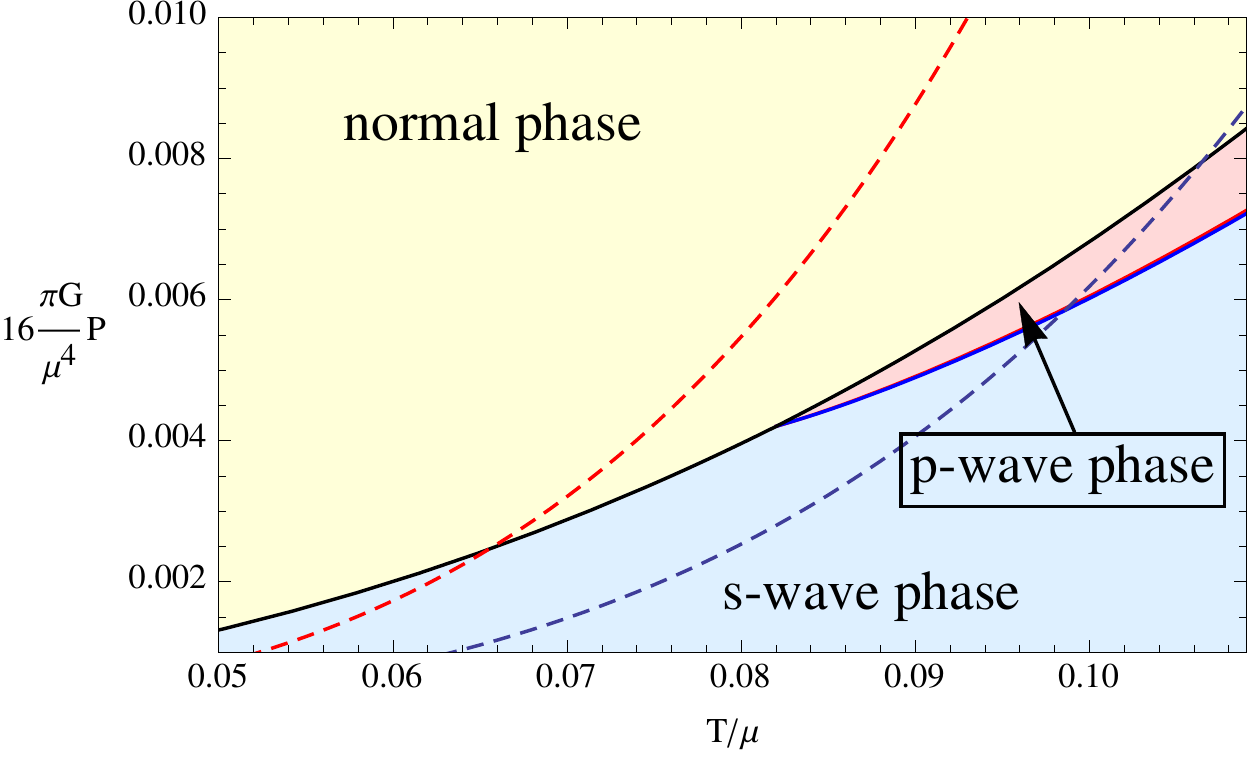}
\includegraphics[width=7.5cm] {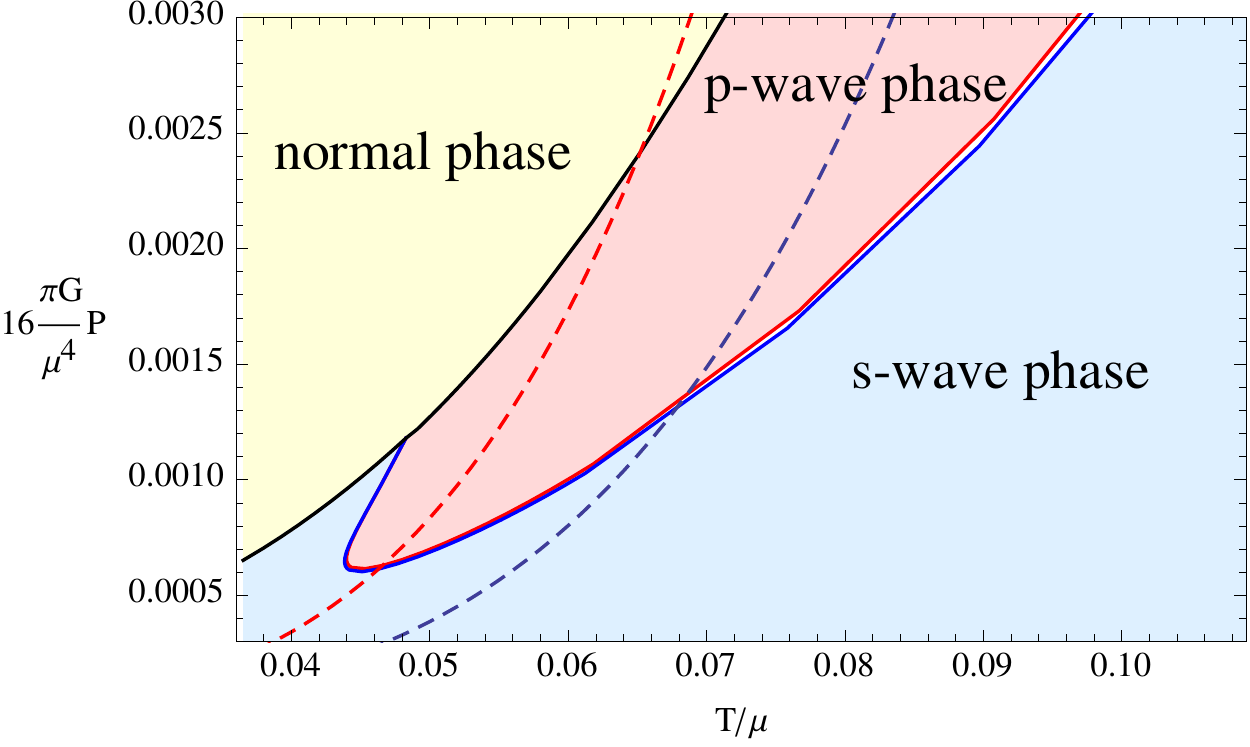}
\caption{\label{PTphasediagram2}Two typical P-T phase diagrams with $\Delta=2.93$(Left) and $\Delta=2.96$(Right). The color codes here are the same as that in Figure~\ref{PTphasediagram}.}
\end{figure}

We use this definition of pressure to construct the P-T phase diagrams with $\Delta=2.93$ and $\Delta=2.96$ for our holographic system in Figure~\ref{PTphasediagram2}. We can see in this figure that the s+p phase between the s-wave phase and the p-wave phase is still too narrow to be seen clearly, this is similar to the results in the first subsection. However, in Figure~\ref{PTphasediagram2}, the normal phase dominate the region with lower temperature and higher pressure, while the superfluid phases are favored in higher temperature and lower pressure region. This result is quite different from that in Figure~\ref{PTphasediagram} as well as the Helium-3 system.

We also draw the two dashed lines to mark the causality bound $\alpha/L^2>-7/36$ (blue) and $\alpha /L^2<9/100$(red). We can see that the curves with constant $\alpha$ are not horizontal lines as in Figure~\ref{PTphasediagram}. This is because that the pressure in the conformal fluid depends on the temperature. We can substitute (\ref{Temperature}) into the formula of pressure (\ref{pressure2}) and get
\begin{eqnarray}\label{pressure2ofT}
\frac{16\pi G}{\mu^4}P=(\frac{T}{\mu})^4\frac{\pi^4 L^6}{L_{\text{eff}}^3}=(\frac{T}{\mu})^4\frac{8 \pi^4 L_{\text{eff}}^3}{(1+\sqrt{1-4\bar{\alpha}}) ^3}.
\end{eqnarray}
We can see from (\ref{pressure2ofT}) that the pressure is proportional to $T^4$, as a result, the constant $\alpha$ curves in the P-T phase diagrams are no longer horizontal. Moreover, we can see at a constant temperature $T$, the pressure is now proportional to $L^6$, and thus increase while $\alpha$ is increasing. Therefore, the red dashed line is now above the blue dashed line, and the superfluid phases are in the high temperature and low pressure region.

\subsection{The two ways to construct P-T phase diagrams}
In the rest of this section, we compare the two different definitions of pressure and the resulting phase diagrams. 

The first definition of the pressure is from the bulk gravitational point of view. When we treat the asymptotically AdS black brane as a thermodynamic system and study it in an extended phase space, we can define this pressure. The dimension of this pressure is also consistent with the one in 4+1 dimensional space time. Therefore in the phase diagrams using this definition of pressure, the phases can be understood as different phases of the bulk black brane. We still use the concepts "normal phase", "s-wave phase", "p-wave phase" and "s+p phase" from the dual field theory, but we should remember that these phases are phases of black brane when we talk about the P-T phase diagrams with this definition of pressure. The phase diagrams are qualitatively similar to that of Helium-3 system, but the physical meanings of the different phases in the gravitational system are vague.

If we want to construct the P-T phase diagrams of the dual field theory, we should define the pressure as in the second subsection. This definition of pressure from energy momentum tensor is the standard one in the study of gauge/gravity duality. Thus in the phase diagrams in Figure~\ref{PTphasediagram2}, we can discuss the different phases in boundary field theory. The phase structure in Figure~\ref{PTphasediagram2} is quite different from that in Figure~\ref{PTphasediagram}. The superfluid phases in Figure~\ref{PTphasediagram} are in the region with lower temperature and higher pressure, while these superfluid phases in Figure~\ref{PTphasediagram2} are in the region with higher temperature and lower pressure.

\section{\bf Conclusions and discussions}\label{sect:conclusion}
In this paper, we studied the holographic s+p model in 5-dimensional Einstein-Gauss-Bonnet gravity in probe limit. We found that the phase transition behaviors and $\Delta-T$ phase diagrams in different values of Gauss-Bonnet coefficient are qualitatively the same as the results in 4-dimensional Einstein gravity. The effect of Gauss-Bonnet term on the phase diagram is rather simple. When the Gauss-Bonnet coefficient is larger, the critical temperatures of phase transitions from the normal phase are lower, and the the value of $\Delta$ for the quadruple intersection point is larger.

In addition, with $L_{\text{eff}}=1$, we constructed the P-T phase diagrams using two different ways of defining the pressure. One definition is the one used in recent study of the P-V-T criticality of the black hole systems, the other is the standard definition from the energy momentum tensor of the boundary field theory. We found that with the first definition, the P-T phase diagrams of this holographic model share similar properties to that of the liquid Helium-3 system, but this definition of pressure is only well defined in the gravitational system. With the second definition of pressure, we can get P-T phase diagrams for the boundary field theory, but the structure of the phase diagrams are quite different from the diagrams with the first definition of pressure.

In this paper, we focused on the phase structure of the system, and haven't drawn figures to show the condensation behavior and free energies. But one can still get the qualitative condensation behavior at fixed values of $\Delta$ and $\alpha$ from the horizontal lines in the phase diagrams. The condensation behaviors of s-wave and p-wave orders are the same as that from the 4-dimensional bulk space time in probe limit, as was shown in Ref.~\cite{Nie:2013sda}. We have also checked the free energy of the different cases, and the s+p phase is always stable.

There are still many limitations in this study. The most obvious one is that in order to get the P-T phase diagrams, we use Gauss-Bonnet coefficient to tune the pressure while fixing the value of $L_{\text{eff}}$. Therefore, the causality constrain on Gauss-Bonnet coefficient also impose a strange constrain in the P-T phase diagrams. More over, the Gauss-Bonnet coefficient is a model parameter rather than a state parameter, thus the way we get the P-T phase diagram may be not quite rigorous. Another limitation is that we only work in probe limit, where the background metric is not affected by the condensation of matter fields. As a result, the pressure of the system is always isotropic in any of the phases including the p-wave one.

To solve the above problems, we need to construct the P-V phase diagram with considering the back reaction of the matter fields on the metric. In that case, we can work in extended phase space to show the P-V-T critical behavior of the holographic s+p model. Moreover, we can study the anisotropy of the pressure in the p-wave and s+p phases. We can also compare the P-T phase diagrams from different methods and get more clues. Finally, we can try to combine the study on the superfluid phase transitions together with the liquid-gas phase transition in holographic models from asymptotically AdS blackholes with spherical horizon, there we might get a more complete P-T phase diagram unifying both the superfluid phase transitions and the liquid gas phase transition. These work are left for our future study.

\acknowledgments
ZYN would like to thank Rong-Gen Cai for useful suggestions on the modified edition, and thank Fedor Kusmartsev, Qi-Yuan Pan, Julian Sonner, Run-Qiu Yang, Jan Zaanen for useful discussions and comments, and thank the organizers of ``Quantum Gravity, Black Holes and Strings", the organizers of ``International Workshop on Condensed Matter Physics $\&$ AdS/CFT" and the organizers of ``Holographic duality for condensed matter physics" for their hospitality. This work was supported in part by the Open Project Program of State Key Laboratory of Theoretical Physics, Institute of Theoretical Physics, Chinese Academy of Sciences, China (No.Y5KF161CJ1), in part by two Talent-Development Funds from Kunming University of Science and Technology under Grant Nos. KKZ3201307020 and KKSY201307037, and in part by the National Natural Science Foundation of China under Grant Nos.  11035008, 11247017, 11375247, 11447131 and 11491240167.


\end{document}